\documentclass[letterpaper, 10 pt, conference]{ieeeconf}  

\IEEEoverridecommandlockouts                              

\usepackage{cite}
\usepackage{amsmath,amssymb,amsfonts}
\usepackage{algorithmic}
\usepackage{graphicx}
\usepackage{textcomp}
\usepackage{balance}
\usepackage{soul,color}
\usepackage[table]{xcolor}
\definecolor{perfblue}{HTML}{DCEAF7}   
\definecolor{effgreen}{HTML}{DFF0DF}  
\definecolor{speedorange}{HTML}{FCE6D5} 
\def\BibTeX{{\rm B\kern-.05em{\sc i\kern-.025em b}\kern-.08em
    T\kern-.1667em\lower.7ex\hbox{E}\kern-.125emX}}

\title{\LARGE \bf
Performance and Energy Trade-Off Analysis of Hierarchical Federated Learning for Plant Disease Classification
}

\author{Athanasios~Papanikolaou\textsuperscript{\ddag},~Athanasios~Tziouvaras\textsuperscript{†},~Pavlos~Stoikos\textsuperscript{†},~Apostolos~Xenakis\textsuperscript{†},\\~Shameem~A~Puthiya Parambath\textsuperscript{+}, George~Floros\textsuperscript{\S}, Enrica~Zereik,\textsuperscript{*} Ivan~Petrovic,\textsuperscript{\ddag}  and~Fabio~Bonsignorio\textsuperscript{\ddag}\\
\textsuperscript{\ddag}University of Zagreb, Croatia, emails:
\{athanasios.papanikolaou, ivan.petrovic, fabio.bonsignorio\}@fer.unizg.hr\\
\textsuperscript{†}University of Thessaly, Greece, emails:
\{attziouv, pastoikos, axenakis\}@uth.gr\\
\textsuperscript{+}University of Glasgow, UK, email:
sham.puthiya@glasgow.ac.uk\\
\textsuperscript{\S}Trinity College Dublin, Ireland, email:
florosg@tcd.ie\\
\textsuperscript{*}Italian National Research Council, Institute of Marine Engineering, Italy, email:
enrica.zereik@cnr.it
}

\begin{document}

\maketitle

\thispagestyle{empty}
\pagestyle{empty}

\begin{abstract}

Early detection of plant diseases is critical for improving crop productivity, while it also facilitates the foundations of precision agriculture. Recent advances in distributed deep learning have enabled plant disease classification models to be trained across geographically distributed agricultural sensing infrastructures. However, deploying such systems in large-scale Internet of Things (IoT) environments, introduces significant challenges related to computational cost, energy consumption, and system efficiency. 
In this paper, we present a design-space exploration of hierarchical federated learning architectures for plant disease classification, with a particular focus on the trade-offs between predictive performance and energy efficiency. We further introduce a power- and energy-aware optimization framework that enables the systematic evaluation and selection of model–aggregator configurations under varying deployment constraints. The hierarchical federated architecture organizes distributed clients through intermediate aggregation layers, reducing communication and computational overhead.
We evaluate multiple convolutional neural network architectures, including EfficientNet-B0, ResNet-50, and MobileNetV3-Large, in combination with different federated aggregation strategies such as FedAvg, FedProx, and FedAvgM. Experimental results demonstrate that different model–aggregator combinations exhibit distinct performance–energy trade-offs. Consequently, we highlight configurations that achieve competitive diagnostic accuracy and significantly reduce system resource requirements.

\end{abstract}

\section{Introduction}
Agriculture plays a critical role in global food security, yet it remains highly vulnerable to plant diseases that can significantly reduce crop yield and quality \cite{WOLFERT201769}. Early and accurate detection of plant diseases is essential for timely interventions and  minimization of economic losses \cite{buja2021advances}. In recent years, the integration of Deep Learning (DL) and Internet of Things (IoT) technologies has transformed precision agriculture by introducing automated, data-driven monitoring of crop health. In particular, image-based DL models have demonstrated high effectiveness in the identification of disease patterns directly from leaf images, while IoT sensor networks allow continuous data acquisition in real-world field environments \cite{9761267}.

Despite these advances, most DL-based systems rely on centralized training and inference paradigms, where data must be transmitted to cloud infrastructures for processing \cite{joshi2023enabling}. This approach introduces significant challenges in IoT environments, particularly in terms of communication overhead, latency, and energy consumption. Continuous data transmission from distributed sensing devices to centralized servers increases network traffic and leads to delays that can hinder real-time decision-making in agricultural applications \cite{mowla2023internet}. At the same time, the computational demands of training and deploying deep neural networks impose substantial energy requirements, which are often incompatible with the limited power budgets of edge devices \cite{TARIQ2025107342}.

These limitations become even more critical in large-scale deployments, where numerous heterogeneous IoT nodes participate in the learning process \cite{8333700}. In such settings, both local computation and communication contribute significantly to the overall system cost, making the selection of efficient model architectures and training strategies essential \cite{zhu2021federated}. Moreover, different combinations of deep learning models and aggregation mechanisms can lead to markedly different performance and energy profiles, further complicating system design. As a result, there is a growing need for approaches that explicitly account for the trade-offs between predictive performance, execution cost, and energy consumption in distributed learning environments.

\begin{figure*}[t]
    \vspace{0.2cm}
    \centering
    \includegraphics{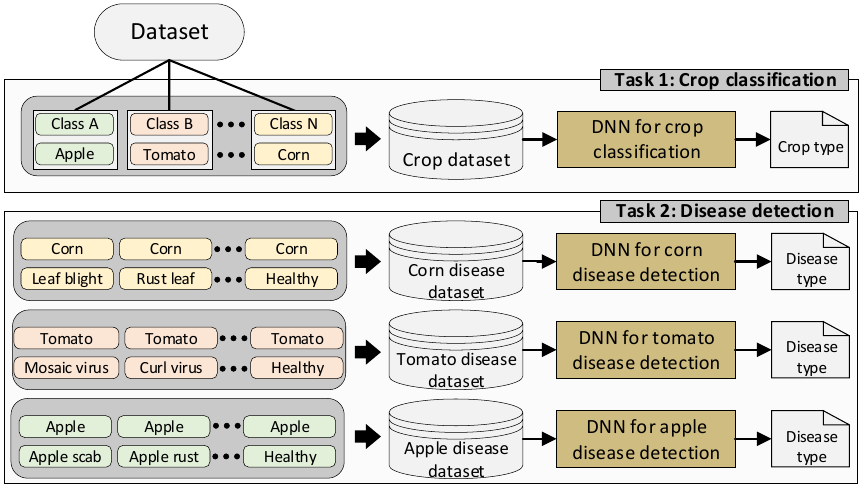}
    \caption {The hierarchical FL concept, where a problem is broken down into concrete subtasks, each one solvable by a DNN model.}
    \label{fig:hier_fl}
\vspace{-0.5cm}
\end{figure*}

To this end, in this paper we address these challenges through a systematic analysis and optimization-driven perspective. The main contributions of this paper are summarized as follows. \textit{First}, we perform a comprehensive design-space exploration of hierarchical federated learning architectures for plant disease classification, analyzing the combined impact of deep learning models and aggregation strategies. \textit{Second}, we introduce a power- and energy-aware optimization framework that enables the joint evaluation of predictive performance, total energy consumption, and total execution time, supporting informed configuration selection under deployment constraints. \textit{Finally}, we evaluate multiple state-of-the-art convolutional neural networks, including EfficientNet-B0, ResNet-50, and MobileNetV3, in combination with federated aggregation methods such as FedAvg, FedProx, and FedAvgM. 

The remainder of this paper is organized as follows: Section \ref{back} presents the background, including an overview of hierarchical federated learning and the deep neural network models considered. Section \ref{power} introduces the proposed power- and energy-aware optimization framework, detailing the problem formulation and the associated optimization strategies. Section \ref{exp} describes the experimental setup and presents the evaluation results. Finally, Section \ref{conl} concludes this work.

\section{Background} \label{back}
\subsection{Hierarchical federated Learning}

In general, Federated Learning (FL) refers to a distributed learning paradigm where a number of individual Deep Neural Network (DNN) models are locally trained and then aggregated together into a global model. Under the FL premise, we consider several devices (also mentioned as nodes) which are spread within a geographical area and collect data from their environment. Each device partially trains a DNN model to fit local data and then it dispatches the partially trained model to a global node. This node aggregates all the collected local models into one unified global model and broadcasts it back to the devices. This process marks the end of one training round, after which the devices continue the training operation using their local datasets.

In this work, we explore the impact of several federated learning methods in terms of performance and total energy consumption. For this reason, we adopt the concept of hierarchical learning to support different model aggregation techniques \cite{b8}. In hierarchical learning, a complex problem is broken down into simpler tasks, which are sequentially solved using a DNN, as illustrated in Figure \ref{fig:hier_fl}. For example, in this work, hierarchical FL decomposes the main task into two subtasks: \textbf{(i)} First identify the plant type (e.g. apple, tomato, corn etc.) using a DNN; and then \textbf{(ii)} assess the disease (if any) of the corresponding plant. Evidently, each plant type has its own set of potential diseases; thus, a dedicated DNN is trained to identify them.

In this context, we perform a design space exploration for different model aggregation functions in order to assess the performance-energy trade-offs within the FL training operation. More specifically we employ the following model aggregation techniques:
\textbf{(i)} \textit{fedavg} \cite{b12}, which averages the weights of the locally trained models to create the global model; \textbf{(ii)} \textit{fedprox} \cite{b13}, which is specifically designed to addresses the challenges of heterogeneous data and tries to stabilize the training process by utilizing information regarding the computational capacities of the participating devices; and \textbf{(iii)} \textit{fedavgm} \cite{b14}, that applies an adaptive learning rate mechanism at the global level, when combining the partially trained models received from individual devices.

\subsection{DNN models}

In order to properly explore the effects of different model aggregation techniques, we also evaluate how each method affects different DNN models. For this exploration, we opt to use the following state of the art models, which are tailored for image classification tasks:
\textbf{(i)} \textit{EfficientNet-B0} \cite{b9}, a lightweight convolutional neural network (CNN) widely used for image classification tasks; \textbf{(ii)} \textit{ResNet-50} \cite{b10}, again a CNN that introduces new techniques to alleviate the gradient vanishing problem and \textbf{(iii)} \textit{MobileNetV3-Large} \cite{b11}, a lightweight DNN, utilizing novel techniques such as linear bottleneck layers and inverted residual structures, in such a way to reduce the computational complexity of the training process.

\section{Power and Energy-Aware Optimization Framework} \label{power}
In this section, we formulate the selection of a model–aggregator configuration as a power- and energy-aware optimization problem. Instead of evaluating performance and resource consumption independently, we introduce a unified framework that jointly considers classification effectiveness (e.g., F1-score, accuracy) and system efficiency (e.g., energy consumption, execution time). This approach enables a systematic comparison of configurations and supports informed decision-making under diverse operational requirements, including energy-constrained, latency-sensitive, and performance-driven deployment scenarios.

\subsection{Problem Formulation}

Let $c \in \mathcal{C}$ denote a candidate configuration, where $\mathcal{C}$ represents the set of all evaluated combinations of deep learning architectures and federated aggregation strategies. Each configuration corresponds to a specific pairing of a model (e.g., ResNet-50, EfficientNet-B0, MobileNetV3-Large) with an aggregation method (e.g., FedAvg, FedProx or FedAvgM), and is associated with both performance-related and system-level metrics derived from the experimental evaluation.

For each configuration $c$, we define the following quantities. $\mathrm{F1}(c)$ and $\mathrm{Acc}(c)$, representing classification performance, as well as $E(c)$ and $T(c)$, denoting total energy consumption and total execution time, respectively. Since these metrics operate on different numerical scales, we apply min--max normalization to map all values to the range $[0,1]$, enabling direct comparison and combination within a unified objective function. The normalization is defined as:

\begin{equation}
\tilde{x}(c) = \frac{x(c) - x_{\min}}{x_{\max} - x_{\min}}
\end{equation}
where $x(c)$ denotes the value of a given metric for configuration $c$, and $x_{\min}$ and $x_{\max}$ correspond to the minimum and maximum values of that metric across all configurations.

To jointly capture the trade-offs between performance and efficiency, we define the following scalar objective:
\begin{equation}
\label{eq:2}
\mathcal{L}(c) = \lambda_1 \,\tilde{E}(c) + \lambda_2 \,\tilde{T}(c) + \lambda_3 \,(1 - \widetilde{\mathrm{F1}}(c))
\end{equation}
where $\tilde{E}(c)$, $\tilde{T}(c)$, and $\widetilde{\mathrm{F1}}(c)$ denote the normalized metrics, and $\lambda_1, \lambda_2, \lambda_3 \geq 0$ are weighting coefficients that control the relative importance of energy consumption, latency, and predictive performance.

The optimization problem is then expressed as:
\begin{equation}
c^* = \arg\min_{c \in \mathcal{C}} \mathcal{L}(c)
\end{equation}

This formulation provides flexibility to adapt the selection process to different deployment scenarios. For instance, increasing $\lambda_1$ prioritizes energy efficiency, while higher values of $\lambda_3$ emphasize predictive performance. In this way, the proposed framework enables a configurable mechanism for selecting the most suitable model--aggregator pair based on system-level constraints and application requirements.

\subsection{Constrained Optimization Formulation}

In addition to the weighted objective formulation, the configuration selection problem can be expressed in a constrained optimization form that reflects practical system limitations. In many real-world deployments, particularly in edge and IoT environments, energy consumption and execution time are subject to strict upper bounds due to hardware and operational constraints.

Under this perspective, the goal is to maximize predictive performance while ensuring that resource usage remains within predefined limits. This can be formulated as:

\begin{equation}
\max_{c \in \mathcal{C}} \mathrm{F1}(c)
\quad \text{s.t.} \quad
E(c) \leq E_{\max}, \;\;
T(c) \leq T_{\max}
\end{equation}

where $E_{\max}$ and $T_{\max}$ denote the maximum allowable energy consumption and execution time, respectively. This formulation enables the identification of configurations that satisfy system-level constraints while achieving the highest possible classification performance.

The constrained formulation is particularly relevant for agricultural scenarios where resource budgets are fixed and must be strictly enforced. By adjusting the values of $E_{\max}$ and $T_{\max}$, the framework can accommodate a wide range of operational conditions, from highly energy-constrained edge devices to more flexible cloud-based environments.

Furthermore, the proposed approach complements the weighted objective formulation by providing an alternative perspective in which constraints are explicitly enforced rather than implicitly incorporated into a scalar objective. As a result, it enhances the interpretability of the selection process and facilitates practical decision-making in resource-aware federated learning systems.

\subsection{Energy Efficiency Metric} \label{sec:eta}

To provide an intuitive and easily interpretable measure of the trade-off between predictive performance and energy consumption, we define an energy efficiency metric that captures the amount of classification performance achieved per unit of energy. Specifically, for each configuration $c \in \mathcal{C}$, the energy efficiency is defined as:

\begin{equation}
\label{eq:5}
\eta(c) = \frac{\mathrm{F1}(c)}{E(c)}
\end{equation}

where $\mathrm{F1}(c)$ denotes the F1-score and $E(c)$ represents the total energy consumption. Higher values of $\eta(c)$ indicate more favorable configurations, as they achieve higher predictive performance while consuming less energy.

This metric is particularly useful in energy-constrained environments, where maximizing performance alone may not be sufficient. By incorporating both accuracy-related and resource-related aspects into a single ratio, $\eta(c)$ enables straightforward ranking and comparison of different model--aggregator combinations.

Although simplified, the energy efficiency metric provides a practical and effective criterion for identifying configurations that balance performance and energy consumption, complementing the optimization formulations presented in the previous subsections.

\section{Experimental Results} \label{exp}

\subsection{Experimental Setup}

To investigate the design space of hierarchical federated plant disease classification, we consider a simulated FL environment in which multiple devices collaboratively train image classification models without exchanging raw data. Each client performs local optimization on its private dataset and, after a fixed number of local epochs, transmits the updated model parameters to a central server. The server aggregates the client updates and broadcasts the resulting global model to initiate the next communication round. This process is repeated for all examined backbone--aggregator configurations and provides a controlled setting for comparing predictive performance, convergence behavior, execution cost, and energy consumption.

The experiments are conducted on the PlantDoc dataset \cite{dataset}, which contains 230,701 RGB leaf images annotated with 38 crop--disease classes spanning 14 crop categories. Since each label jointly encodes the crop identity and its health condition, we formulate the task hierarchically rather than as a single flat classification problem. More specifically, a crop classifier is first trained to identify the crop category, after which a crop-specific disease classifier is used to determine the disease state within the predicted crop. This decomposition is intended to reduce inter-crop ambiguity, while enabling the disease models to specialize on a smaller and semantically consistent subset of classes.

To more closely reflect real-world field conditions and assess model robustness under practical sources of variation, such as illumination, viewpoint, focus, compression, and framing, we employed the data-augmented dataset presented in our previous work \cite{b8}. The augmentation procedure is reported in detail in that study. In brief, the images of each class were split into five equally sized subsets, and a distinct augmentation recipe was applied to each subset. This strategy maintains the original number of images per class while keeping the five augmentation variants evenly distributed, with each representing $20\%$ of the class data. Following this protocol, we define five Use Cases (UCs) for the comparative experiments:

\begin{itemize}
\item \textbf{UC1:} \textit{SunnyAngle}, simulating bright sunlight, oblique viewpoints, mild perspective changes, rotation, contrast enhancement, and soft shadows.
\item \textbf{UC2:} \textit{OvercastNoise}, simulating darker, lower-contrast, slightly desaturated images with Gaussian sensor noise.
\item \textbf{UC3:} \textit{Defocus}, simulating mild blur, zoom jitter, small rotations, motion blur, wind-driven motion, or shallow depth-of-field misfocus.
\item \textbf{UC4:} \textit{JPEGandCast}, simulating JPEG recompression artifacts and warm or cool color shifts caused by compression pipelines or white-balance drift.
\item \textbf{UC5:} \textit{OffCenter}, simulating imperfect framing through off-center cropping, re-centering, and light exposure or contrast changes.
\end{itemize}

The training procedure follows the same hierarchical logic. The crop classifier is trained using the complete dataset, since every image contributes to crop recognition. Subsequently, a separate federated training session is performed for each crop-specific disease classifier using only the samples that belong to the corresponding crop. As a result, a client participates in a given crop-specific session only if its local partition contains samples from that crop. This design avoids irrelevant updates across unrelated crops and preserves distinct training states for the routing stage and the downstream disease classifiers. During inference, the pipeline operates sequentially: the crop classifier first determines the crop category, and the corresponding disease classifier is then invoked to produce the final prediction.

\begin{table}[t]
\vspace{0.2cm}
\centering
\caption{The experimental setup used through the experimentation process.}
\label{tab:exp_setup}
\renewcommand{\arraystretch}{1.05}
\setlength{\tabcolsep}{3pt}
\footnotesize
\begin{tabular}{p{0.36\linewidth} p{0.56\linewidth}}
\hline
\textbf{Setting} & \textbf{Configuration} \\
\hline
Dataset & PlantDoc, 230,701 images \\
Hierarchy & 14 crops, 38 crop--disease classes \\
Split & 70/15/15 (train/val/test) \\
Partitioning & Deterministic i.i.d., stratified, disjoint \\
Clients / rounds / local epochs & 10 / 30 / 5 \\
Backbones & EfficientNet-B0, ResNet-50, MobileNetV3-Large \\
Strategies & FedAvg, FedProx, FedAvgM \\
Batch size / input size & 64 / $224 \times 224$ \\
Optimizer / LR / WD & Adam / $10^{-4}$ / $10^{-4}$ \\
Loss & Cross-entropy + label smoothing ($0.1$) \\
FedProx & $\mu = 0.01$ \\
FedAvgM & Server LR $=0.3$, momentum $=0.3$ \\
Initialization & ImageNet pretrained \\
Augmentation & Five controlled image degradation settings applied to the validation and testing set \\Logged metrics & Loss, Accuracy, Precision, Recall, F1, total execution time, total energy consumption \\
Hardware & NVIDIA RTX 6000 Ada Generation \\
\hline
\end{tabular}
\vspace{-0.5cm}
\end{table}

For reproducible benchmarking, the dataset is partitioned across clients using an i.i.d. (independent and identically distributed), disjoint, and approximately balanced distribution scheme, so that each client receives a comparable local subset with representation from all classes. In addition to the standard train/validation/test split, the evaluation protocol includes five controlled image degradation settings inherited from the earlier version of the framework, allowing model robustness to be assessed under more challenging visual conditions. Within this setup, we evaluate three backbone architectures---EfficientNet-B0, ResNet-50, and MobileNetV3-Large---combined with the FedAvg, FedProx, and FedAvgM aggregation strategies. The complete training and evaluation protocol, including the main hyperparameters and hardware configuration, is summarized in Table~\ref{tab:exp_setup}, while the resulting predictive metrics, total energy consumption, and total execution time are reported in Table~\ref{tab:all_metrics}; these system-level measurements correspond to the complete experiment for each backbone--aggregator configuration, including the full hierarchical training procedure over all communication rounds and local epochs, rather than to a single epoch, round, or crop-specific classifier. Finally, in our experimental evaluation, we consider equal weighting coefficients $\lambda_1 = \lambda_2 = \lambda_3 = \frac{1}{3}$, corresponding to a balanced trade-off between predictive performance, total energy consumption, and total execution time. However, rather than explicitly computing a single scalar objective value, we report the individual metrics to provide a more transparent comparison across configurations.

\subsection{Experimental Evaluation}

To keep the scope of the present study focused, we do not further discuss the plant classification module here. That stage has already been described in our previous work \cite{b8}, and it is not the primary subject of this paper. The experimental analysis in this section is specifically aimed at comparing the candidate backbone--aggregator configurations within the federated disease-classification pipeline. The plant classifier is external to this comparison, and the reported experiments were conducted independently of that additional stage. Although its inclusion in a complete end-to-end deployment would introduce a small additional computational cost and a minor decrease in overall predictive performance, it would affect all configurations in a similar manner and therefore does not alter the comparative conclusions of this study.

\begin{figure*}[t]
    \vspace{0.2cm}
    \centering
    \includegraphics[width=\textwidth]{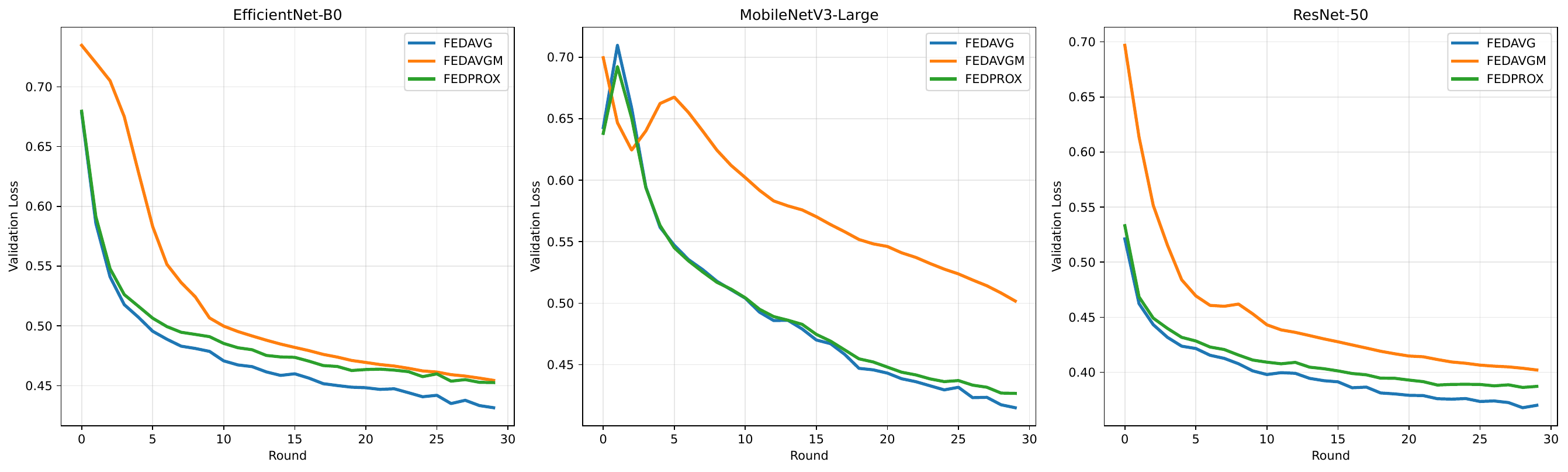}
    \caption {EfficientNet-B0/ResNet-50/MobileNetV3-Large loss over epochs, using fedavg, fedprox and fedavgm.}
    \label{fig:all_models_loss}
\end{figure*}

\begin{table*}[h]
\caption{Average predictive performance, total energy consumption, total execution time, and energy-efficiency score ($\eta$) for each complete model--aggregator experiment. Blue highlights denote the strongest predictive configuration, green highlights denote the configuration that achieves both the lowest total energy consumption and the highest energy-efficiency score, and orange highlights denote the fastest configuration.}
\label{tab:all_metrics}
\centering
\resizebox{0.92\textwidth}{!}{
\begin{tabular}{c|c|ccccccc}
Model & Aggregator & Accuracy & Recall & Precision & F1-score & Total energy (Wh) & Total execution time (s) & $\eta$ \\
\hline
EfficientNet-B0 & fedavg  & 0.8632 & 0.8680 & 0.8845 & 0.8535 & 163.39 & 4216.38 & 0.005 \\
EfficientNet-B0 & fedprox & 0.8519 & 0.8579 & 0.8787 & 0.8429 & 218.76 & 5503.40 & 0.004 \\
EfficientNet-B0 & fedavgm & 0.8464 & 0.8515 & 0.8780 & 0.8366 & 257.24 & 5121.62 & 0.003 \\
\hline
\cellcolor{perfblue} ResNet-50 & \cellcolor{perfblue} fedavg  & \cellcolor{perfblue} 0.9169 & \cellcolor{perfblue} 0.9093 & \cellcolor{perfblue} 0.9228 & \cellcolor{perfblue} 0.9062 & 575.12 & 13315.68 & 0.002 \\
ResNet-50 & fedprox & 0.9067 & 0.8968 & 0.9149 & 0.8942 & 372.12 & 12291.58 & 0.002 \\
ResNet-50 & fedavgm & 0.8987 & 0.8890 & 0.9060 & 0.8840 & 461.02 & 12147.10 & 0.002 \\
\hline
MobileNetV3-Large & fedavg  & 0.8780 & 0.8817 & 0.8937 & 0.8652 & 176.68 & 5112.66 & 0.005 \\
\cellcolor{effgreen} MobileNetV3-Large & \cellcolor{effgreen} fedprox & 0.8710 & 0.8751 & 0.8910 & 0.8583 & \cellcolor{effgreen} 142.73 & 4446.40 & \cellcolor{effgreen} 0.006 \\
\cellcolor{speedorange} MobileNetV3-Large & \cellcolor{speedorange} fedavgm & 0.8191 & 0.8242 & 0.8674 & 0.7992 & 165.90 & \cellcolor{speedorange} 4202.80 & 0.005 \\
\hline
\end{tabular}
}
\vspace{-0.5cm}
\end{table*}

The results in Table~\ref{tab:all_metrics}, together with the validation-loss trends shown in Figure~\ref{fig:all_models_loss}, indicate that the proposed hierarchical federated pipeline remains stable across all examined configurations, while clear differences emerge in predictive performance and computational cost. Among the evaluated backbones, ResNet-50 consistently achieves the strongest classification results, with the FedAvg configuration obtaining the highest accuracy and F1-score overall. This suggests that, within the hierarchical per-crop setting, the residual architecture provides the highest discriminative capacity for disease recognition.

When the aggregation strategies are compared within each backbone, FedAvg consistently yields the best predictive performance. For EfficientNet-B0, FedAvg outperforms both FedProx and FedAvgM across all classification metrics while also requiring less energy than the other two variants of the same model. A similar trend is observed for ResNet-50, where FedAvg again provides the best predictive performance, although with the highest total energy consumption among all tested configurations. For MobileNetV3-Large, FedAvg remains the most accurate option, but the gap relative to FedProx is small, indicating that a lightweight architecture can retain competitive diagnostic performance under a more resource-conscious aggregation setting.

The combined validation-loss curves in Figure~\ref{fig:all_models_loss} are consistent with the final test metrics. Across all three backbones, FedAvg converges to the lowest validation loss, while FedProx follows closely with slightly weaker but still stable behavior. FedAvgM is the least favorable strategy in this study, showing consistently higher loss and lower final predictive performance, particularly for MobileNetV3-Large. The agreement between loss evolution and final test metrics indicates that the observed performance differences are systematic and not the result of isolated end-of-training fluctuations.

From the resource perspective, the results reveal a clear trade-off between predictive quality, energy demand, and total execution time. ResNet-50 with FedAvg delivers the best overall classification performance, but also incurs the highest energy cost and the longest total execution time. In contrast, MobileNetV3-Large with FedProx achieves the lowest total energy consumption while maintaining competitive predictive performance, making it the strongest low-energy configuration in this study. The fastest total execution time is obtained by MobileNetV3-Large with FedAvgM, indicating that the most time-efficient configuration is not necessarily the most accurate or the most energy-efficient. EfficientNet-B0 occupies an intermediate position, requiring substantially less energy than ResNet-50 but without matching MobileNetV3-Large in overall efficiency or ResNet-50 in raw predictive accuracy.

To summarize this trade-off more explicitly, we use the energy-efficiency metric $\eta$ from equation~\ref{eq:5}, which expresses the ratio between predictive performance and total energy consumption. As shown in Table~\ref{tab:all_metrics}, MobileNetV3-Large with FedProx achieves the highest $\eta$, indicating the most favorable balance between diagnostic quality and energy usage. By contrast, ResNet-50 with FedAvg, despite achieving the highest predictive performance, exhibits a much lower $\eta$ due to its significantly higher energy requirements. These findings reinforce that lightweight architectures combined with suitable aggregation strategies are more appropriate for resource-constrained deployments, whereas higher-capacity models remain preferable when maximizing predictive performance is the primary objective.

Overall, the reported results provide the necessary inputs for the proposed optimization framework, enabling the evaluation of the candidate configurations under different performance and resource constraints and supporting the selection of the most suitable model--aggregator pair for a given deployment scenario. 

\section{Conclusions} \label{conl}

This paper explores the trade-offs between predictive performance and energy consumption in a hierarchical federated learning framework for plant disease classification. The results demonstrate that ResNet-50 combined with FedAvg achieves the highest predictive accuracy, while lighter models, particularly MobileNetV3-Large with FedProx, offer a more favorable balance between diagnostic performance and energy efficiency. By jointly evaluating classification metrics, validation loss, execution time, and total energy consumption, the study provides a comprehensive assessment of system behavior beyond accuracy alone. Overall, the findings highlight that no single model–aggregator configuration is optimal in all scenarios, and that the best choice depends on deployment constraints such as computational resources, latency requirements, and energy budgets.

The work also suggests future directions, particularly the need to explore different weightings of the optimization parameters ($\lambda_1$, $\lambda_2$, $\lambda_3$), which control the importance of energy, latency, and performance. A systematic sensitivity analysis of these parameters could improve the adaptability of the framework and support more context-aware decision-making in real-world, resource-constrained environments.


\section*{Acknowledgment}
This work was funded under the COIN-3D project, which has received funding from the European Union’s Horizon Europe research and innovation program under grant agreement No. 101159667.

\balance
\bibliographystyle{IEEEtran}
\bibliography{refs}

\end{document}